\documentclass[10pt]{article}
\usepackage{epsfig,dsfont}
\usepackage{geometry}
\begin{document}  
\sffamily

\vspace*{1mm}

\begin{center}

{\LARGE
Density of states method for the $\mathds{Z}_3$ spin model}
\vskip10mm
Christof Gattringer and Pascal T\"orek 
\vskip5mm
University Graz, Institute of Physics\\
Universit\"atsplatz 5, 8010 Graz, Austria 
 
\end{center}
\vskip10mm

\begin{abstract}
We apply the density of states approach to the $\mathds{Z}_3$ spin model with a chemical potential $\mu$. For determining the 
density of states we use restricted Monte Carlo simulations on small intervals of the variable for the density. In each interval
we probe the response of the system to the variation of a free parameter in the Boltzmann factor. 
This response is a known function which we fit to the
Monte Carlo data and the parameters of the density are obtained from that fit (functional fit approch; FFA). We evaluate observables 
related to the particle number and the particle number susceptibility, as well as the free energy. We find that for a surprisingly large range of
 $\mu$ the results from the 
FFA agree very well  with the results from a reference simulation in the dual formulation of the 
$\mathds{Z}_3$ spin model which is free of the complex action problem.
\end{abstract}

\vskip5mm

\begin{center}
{\sl To appear in Physics Letters B.}
\end{center}

\vskip10mm

\noindent  
{\bf Introductory comments:} 
\vskip2mm

\noindent
It is well known that  at finite density Monte Carlo simulations of many lattice field theories are plagued by the complex action problem 
(or sign problem).  At non-zero chemical potential $\mu$ the action $S$ acquires an imaginary part and the Boltzmann factor 
$e^{-S}$ cannot be used as a probability weight. Several different strategies to overcome the complex action problem have been proposed
over the years, among them the density of states (DoS) method, which is the topic of this letter.

Originally the DoS method was introduced to lattice field theories in \cite{dosold} and for some recent applications and a critical review 
see, e.g., \cite{dosrecent1,dosrecent2}. The key problem of DoS techniques is that the density of states $\rho$ varies
over many orders of magnitude. When applying DoS to finite-$\mu$ problems the density $\rho$ is integrated over with a highly
oscillating factor and the frequency of the oscillation increases exponentially with $\mu$. Thus the numerical challenge is twofold: The density 
$\rho$ has to be determined over many orders of magnitude and this determination has to be very precise since $\rho$ is probed with a highly 
oscillating factor. 

An interesting new approach to the accuracy problem in the DoS for lattice field theories was presented in \cite{doskurt1}. 
The idea is related to a proposal by Wang and Landau \cite{wanglandau} and divides the variable of the density $\rho$ into small
intervals where restricted Monte Carlo simulations are performed to determine $\rho$ in that interval. This leads to exponential
error suppression \cite{doskurt1} and was shown to allow for a precise evaluation of observables in pure gauge theory and in SU(2) gauge theory
with a Polyakov loop source \cite{doskurt1}. 
Furthermore, in \cite{doskurt2} the density and the complex phase were studied in the $\mathds{Z}_3$ spin model
with chemical potential, which is a simple effective theory for the center degrees of freedom of QCD \cite{z3model}. This model is
particularly interesting since it can be mapped to a dual representation without complex action problem 
where precise Monte Carlo simulations at finite $\mu$ are possible \cite{z3dual}.

In this letter we develop the DoS for the $\mathds{Z}_3$ spin model 
further (functional fit approach; FFA), using again the dual simulation \cite{z3dual} as our reference data.
We evaluate observables related to the quark number density, the quark number susceptibility as well as the free energy 
and show that the corresponding dual results
can be reproduced for a surprisingly large range of the chemical potential. First results with the FFA were presented in \cite{poslat}.

\newpage
\noindent  
{\bf Definition of the model and the density of states:} 
\vskip2mm

\noindent
The $\mathds{Z}_3$ spin model in an external field with strength $\kappa$, a chemical potential $\mu \beta$ (in units of the
underlying QCD inverse temperature $\beta$) and an effective temperature parameter $\tau$ is described by the action
 
\begin{equation}
S[P]  =  \sum_x\left[\tau\sum_{\nu=1}^3 \big(P_x^\star P_{x+\hat{\nu}} + c.c. -2\big) + \kappa  e^{\mu\beta} (P_x-1)  + 
\kappa e^{-\mu \beta} (P_x^\star-1) \right],
\label{action}
\end{equation}
where the dynamical degrees of freedom are the spins $P_x\in \mathds{Z}_3 = \{1,e^{i2\pi /3},e^{-i2\pi /3}\}$, living on the sites
$x$ of a 3-dimensional lattice with periodic boundary conditions. The action is normalized such that $S[P]=0$ if $P_x=1\,\forall x$. 
The partition function is obtained by summing the Boltzmann factor
over all configurations $P$, i.e.,  $Z = \sum_{\{P\}} e^{S[P]}$. The first and second derivatives
of $\ln Z$ with respect to $\mu \beta$ have the interpretation of the quark number density and the quark number susceptibility.
It is obvious that for finite chemical potential, $\mu\beta \neq 0$, 
the action (\ref{action}) has a non-zero imaginary part and the model has a complex action problem.   

We remark at this point that for the $\mathds{Z}_3$ spin model $\mu$ does not play the role of a chemical potential in the strict sense, 
since there is no conserved charge it couples to. Still, we refer to $\mu$ as the ''chemical potential'' due to its origin from the proper chemical 
potential of QCD and due to the fact that it generates the complex action problem in exactly the same way as in QCD. The connection of the 
effective $\mathds{Z}_3$ model to QCD is obtained from a strong coupling expansion (see, e.g., \cite{effmod}) combined with a hopping 
expansion of the fermion determinant. This connection also provides the physical interpretation of the spins and of the parameters: 
The spins $P_x$ correspond to the Polyakov loops of the underlying lattice QCD formulation, restricted to the center values of SU(3), i.e., the group 
$\mathds{Z}_3$. The strong coupling expansion identifies $\tau$ as a parameter which is an increasing function  of the temperature $T$ of the 
underlying lattice QCD theory, while the hopping expansion shows that $\kappa$ is proportional to the number of quark flavors and a function of the 
QCD quark mass $m$ which decreases with increasing $m$. 

For a convenient notation we introduce abbreviations for the total numbers of spins pointing in each of the three possible 
directions as
$N_0[P]  =  \sum_x \delta\left(P_x,1\right)$ and $N_\pm[P]  =  \sum_x\delta\left(P_x,e^{\pm i 2\pi/3}\right)$.
Obviously $N_0[P] + N_+[P] + N_-[P] = V$, where $V$ denotes the lattice volume, i.e., the total number of sites of the lattice.
Using these we split the action into real and imaginary parts, i.e., $S[P] = S_R[P]  +  iS_I[P]$, with 
\begin{equation}
S_R[P] =  \tau \sum_x \sum_{\nu=1}^3\big(P_x^\star P_{x+\hat{\nu}} + c.c.-2\big)  + \kappa \, 3(N_0[P] - V) \cosh (\mu \beta) 
\quad , \quad
S_I[P] =  \kappa \, \sqrt{3} \, \sinh (\mu \beta) \, \Delta N[P] \; , 
\label{eq1.2} 
\end{equation}
where we have defined $\Delta N[P] = N_+[P] - N_-[P] \; \in \; \{-V,-V+1,\dots,V\}$.
The variables $N_+[P]$, $N_-[P]$  and $\Delta N[P]$ behave under complex conjugation of all spin variables $P_x \rightarrow P_x^*$ as
$N_+[P] \rightarrow N_-[P]$, $N_-[P]  \rightarrow N_+[P]$,
$\Delta N[P] \rightarrow - \Delta N[P]$,
and consequently $S[P^\star] \; = \; S_R[P] \, - \, iS_I[P] \; = \; S[P]^\star$. Exploring this relation the partition sum can be written as
\begin{equation}
Z   =  \sum_{\{P\}} e^{S[P]}  = \ \sum_{\{P\}} e^{S_R[P]} \, e^{iS_I[P]}  =  \sum_{\{P\}} e^{S_R[P]} \, \cos\left(S_I[P]\right) =  
\sum\limits_{\{P\}} e^{S_R[P]}\cos\left(\kappa \sqrt{3} \sinh( \mu \beta) \; \Delta N[P] \right)  .
\label{partitionsum}
\end{equation}
We now define a weighted density of states ($\delta$ here denotes a Kronecker-delta),
\begin{equation}
\rho(d) \; = \; \sum_{\{P\}}e^{S_R[P]} \, \delta\left(d - \Delta N[P]\right) \; , \; \; d = -V, -V+1, \, .... \, V -1, V \; .
\label{rhodef}
\end{equation}
Exploring again the symmetry properties of the action 
one trivially finds that $\rho(d)$ is an even function of $d$, i.e., $\rho(-d) \; = \; \rho(d)$.
Using the density of states, the partition sum (\ref{partitionsum}) can be written as
\begin{equation}
Z  \; = \; \sum_{d\, =\, -V}^{V}\rho(d) \, \cos\left( \kappa \sqrt{3} \sinh(\mu\beta) \; d  \right) \; .
\label{zfinal}
\end{equation}
Expectation values of observables $O(\Delta N)$ which are a function of $\Delta N$ are given by
\begin{equation}
\langle O \rangle  \; = \; \frac{1}{Z} \sum_{d \, = \, -V}^{V}  \rho(d) \, \Big[ \cos\left( \!\kappa \sqrt{3} \sinh( \mu\beta)  \, d  \!\right) O_E(d) \; + \; i 
\sin\left( \!\kappa \sqrt{3} \sinh( \mu\beta)  \, d  \! \right) O_O(d) \Big] \, ,
\label{vev}
\end{equation}
where $O_E$ and $O_O$ denote the even and odd parts of $O(\Delta N)$.

The partition sum (\ref{zfinal}) and the expectation values (\ref{vev}) are obtained by summing the density $\rho(d)$ with the factors
$\cos\left(\kappa \sqrt{3} \sinh(\mu\beta) \, d \right)$ and $\sin\left(\kappa \sqrt{3} \sinh(\mu\beta) \, d \right)$. 
While the density $\rho(d)$ is strictly positive, these factor are oscillating with $d$
and the frequency of oscillation increases exponentially with the chemical potential $\mu \beta$ and linearly
with the strength parameter $\kappa$. Thus for larger values of 
$\mu \beta$ (or $\kappa$) the density $\rho(d)$ has to be computed very accurately. This is how the complex action problem manifests 
itself in the density of states approach. 

\vskip3mm    
\noindent  
{\bf Computing the density of states :} 
\vskip2mm

\noindent
For the numerical computation we parameterize the density of states $\rho(d)$  as
\begin{equation}
\rho(d) \; = \; \prod_{j=0}^{|d|} e^{-a_j} \; = \; \exp \bigg( -\sum_{j=0}^{|d|} a_j \bigg) 
\; , \; \; d = -V, -V+1, \, .... \, V -1, V \; ,
\label{rhoparam}
\end{equation}
with real parameters $a_j$. Note that this parameterization is exact in the sense that it contains $V+1$ parameters, 
precisely the number of independent degrees of freedom $\rho(d)$ has (remember that $\rho(d)$ is an even function). 
We also remark, that an overall normalization of $\rho(d)$ can be chosen freely, since it 
cancels in the expectation values (\ref{vev}). Here we choose the normalization $\rho(0) = 1$, which corresponds to 
setting $a_0 = 0$. 
  
For the calculation of the coefficients $a_j$ we define restricted expectation values $\langle \langle O \rangle \rangle_n(\lambda)$,
$n = 0,1, ... \, V-1$,
which depend on a free parameter $\lambda$,
\begin{equation}
\langle\langle O \rangle\rangle_n (\lambda) = \frac{1}{Z_n(\lambda)} \sum_{\{ P \}} 
\theta_n\big(\Delta N[P]\big) \, e^{S_R[P]} \, e^{\, \lambda \, \Delta N[P]} \, O( \Delta N[P] ) , \;
Z_n(\lambda)  =  \sum_{\{ P \}} 
\theta_n\big(\Delta N[P]\big) \, e^{S_R[P]} \,  e^{\, \lambda \, \Delta N[P]}.
\label{vevrestricted}
\end{equation}
Here we have defined 
\begin{equation}
\theta_0(d) \; = \; \left\{ \begin{array}{cc} 
1 \;  &  \mbox{for} \; d =  0, 1 \\
0 \; &  \mbox{otherwise} 
\end{array} \right. \;  \mbox{for} \; n = 0 \quad , \quad
\theta_n(d) \; = \; \left\{ \begin{array}{cc} 
1 \; &  \mbox{for} \;  |d -n| \leq 1 \\
0 \; &  \mbox{otherwise} 
\end{array} \right. \; \mbox{for} \; n = 1,2, ... \, V - 1 \; .
\end{equation}
Varying the parameter $\lambda$ in (\ref{vevrestricted}) probes the density in the interval set by $\theta_n(d)$ and the response 
of the system to changing $\lambda$ can be used to determine the parameters of $\rho(d)$ in that interval.  
In the restricted expectation values (\ref{vevrestricted}) only real and positive weight factors appear, such that they 
can be evaluated with a restricted Monte Carlo strategy which we will 
discuss below. 

\begin{figure}[t]
\centering
\hspace*{-0mm}\includegraphics[height=92mm]{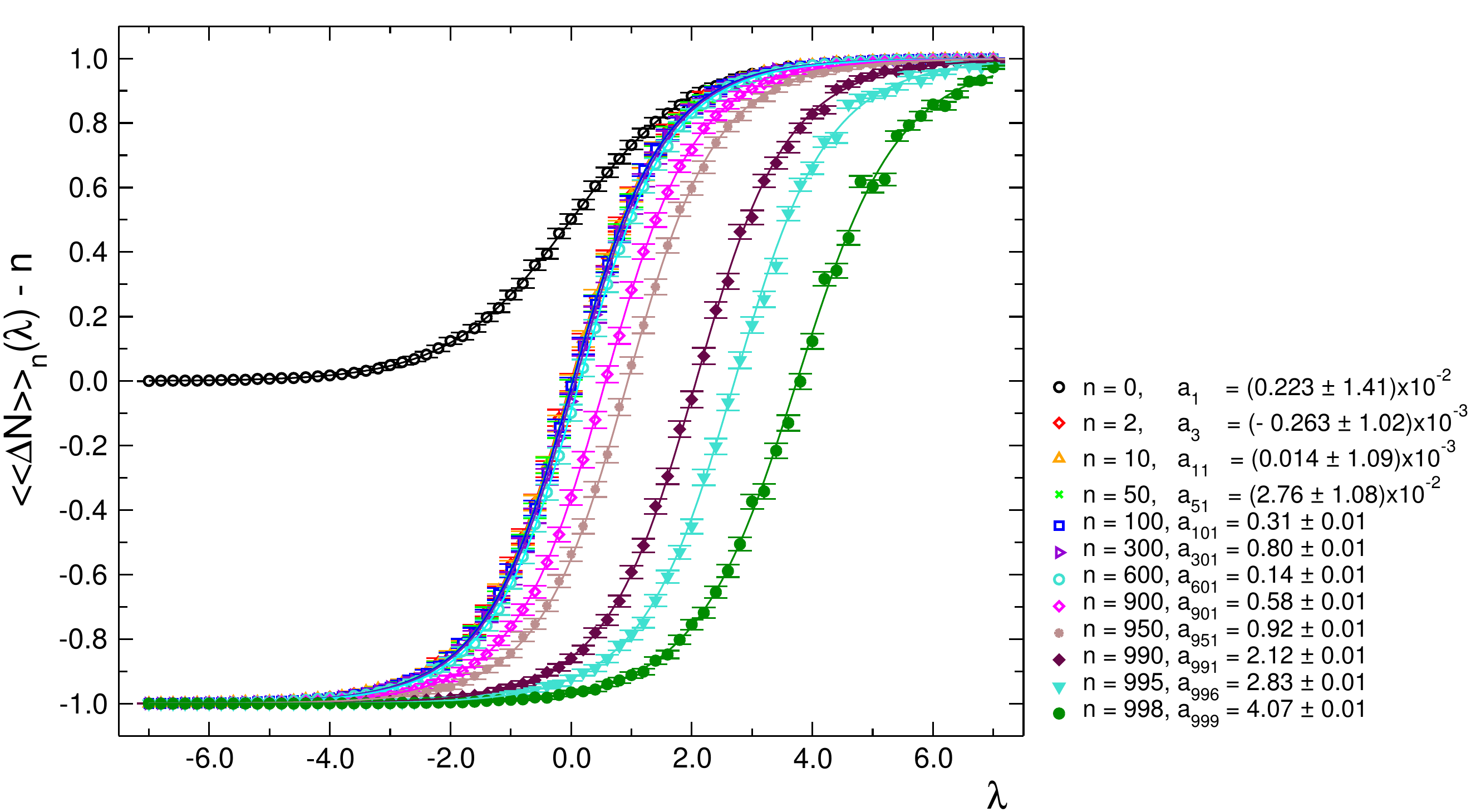}
\caption{Monte Carlo results (symbols) for $\langle \langle \Delta N \rangle \rangle_n(\lambda)\!-\!n$ for 
$n = 0$, 2, 10, 50, 100, 300, 600, 900, 950, 990, 995 and 998 as a function of $\lambda$. The data are for 
a $10^3$ lattice with parameters $\tau = 0.16$, 
$\kappa = 0.01$ and $\mu \beta = 1.0$. The full curves are the fits with the functions on the rhs.\ of (\ref{vevn}).
The resulting values for the corresponding coefficients $a_j$ are given in the legend.}
\label{fig_coeffs}	
\end{figure}

In particular we are here interested in the observable $O = \Delta N$, and now 
use the density of states $\rho(d)$ in the form of (\ref{rhoparam}) to evaluate the restricted  
expectation values $\langle\langle \Delta N \rangle\rangle_n (\lambda)$. A straightforward calculation gives  
\begin{equation}
\langle\langle \Delta N \rangle\rangle_0 (\lambda)  =   
\frac{e^{\lambda - a_1}}
{e^{\lambda - a_1} + 1}  \quad , \quad
\langle\langle \Delta N \rangle\rangle_n (\lambda) - n   = \frac{ e^{2 \lambda - a_n - a_{n+1} } - 1}
{ e^{2 \lambda - a_n - a_{n+1} } +
e^{ \lambda - a_n } + 1 } \; \; , \; \; n = 1, ... V \! - \! 1 \; .  
\label{vevn} 
\end{equation}
The right hand sides are simple functions of $\lambda$: They are monotonically increasing (the derivatives 
with respect to $\lambda$ are easily shown to be positive) and for $n \geq 1$ have a single zero ($\lim_{\lambda \rightarrow -\infty}$ 
is negative, $\lim_{\lambda \rightarrow + \infty}$ is positive). Examples for different $n$ are shown in Fig.~\ref{fig_coeffs}.

Using Monte Carlo simulations we can evaluate  
$\langle\langle \Delta N \rangle\rangle_n (\lambda) - n$ for different values $\lambda_i, \,  i = 1,2, ... \, N_\lambda$ (typically 
$N_\lambda = {\cal O}(10)$) and fit the results according to the right hand sides of (\ref{vevn}). The one-parameter fit 
for $\langle\langle \Delta N \rangle\rangle_0 (\lambda)$ determines the first non-trivial coefficient $a_1$ 
(remember that we chose the normalization $a_0 = 0$). The fit value for $a_1$ can then be inserted in the right hand side 
of (\ref{vevn}) for $n=1$ such that with 
another one-parameter fit of  $\langle\langle \Delta N \rangle\rangle_1 (\lambda) - 1$ we can determine $a_2$, which in turn is then 
inserted in the fit function for $\langle\langle \Delta N \rangle\rangle_2 (\lambda) - 2$, which gives $a_3$ from a one-parameter fit, 
et cetera. Using this sequence of fits we can determine all coefficients $a_j$ from fits of the Monte Carlo data with 
simple functions that depend on only a single parameter (compare Fig.~\ref{fig_coeffs}). Thus we refer to our approach sketched in this letter 
as ''functional fit approach'' (FFA).

We expect that this method has smaller statistical errors for the same numerical effort than iteration methods such as the LLR
\cite{doskurt1,doskurt2} or some general root finding procedure for the $\langle\langle \Delta N \rangle\rangle_n (\lambda) - n$ 
(which also provides iterative equations for the $a_j$). The advantage of the FFA 
comes from the fact that all Monte Carlo data, i.e., the results
for $\langle\langle \Delta N \rangle\rangle_n (\lambda_i) - n$ at all  values $\lambda_i$ of the parameter $\lambda$, are 
used to determine the coefficients $a_j$. Furthermore, it can be seen that possible instabilities from the statistical errors of the Monte Carlo 
data are minimized here.

\begin{figure}[t]
\centering
\hspace*{-1mm}
\includegraphics[height=56mm]{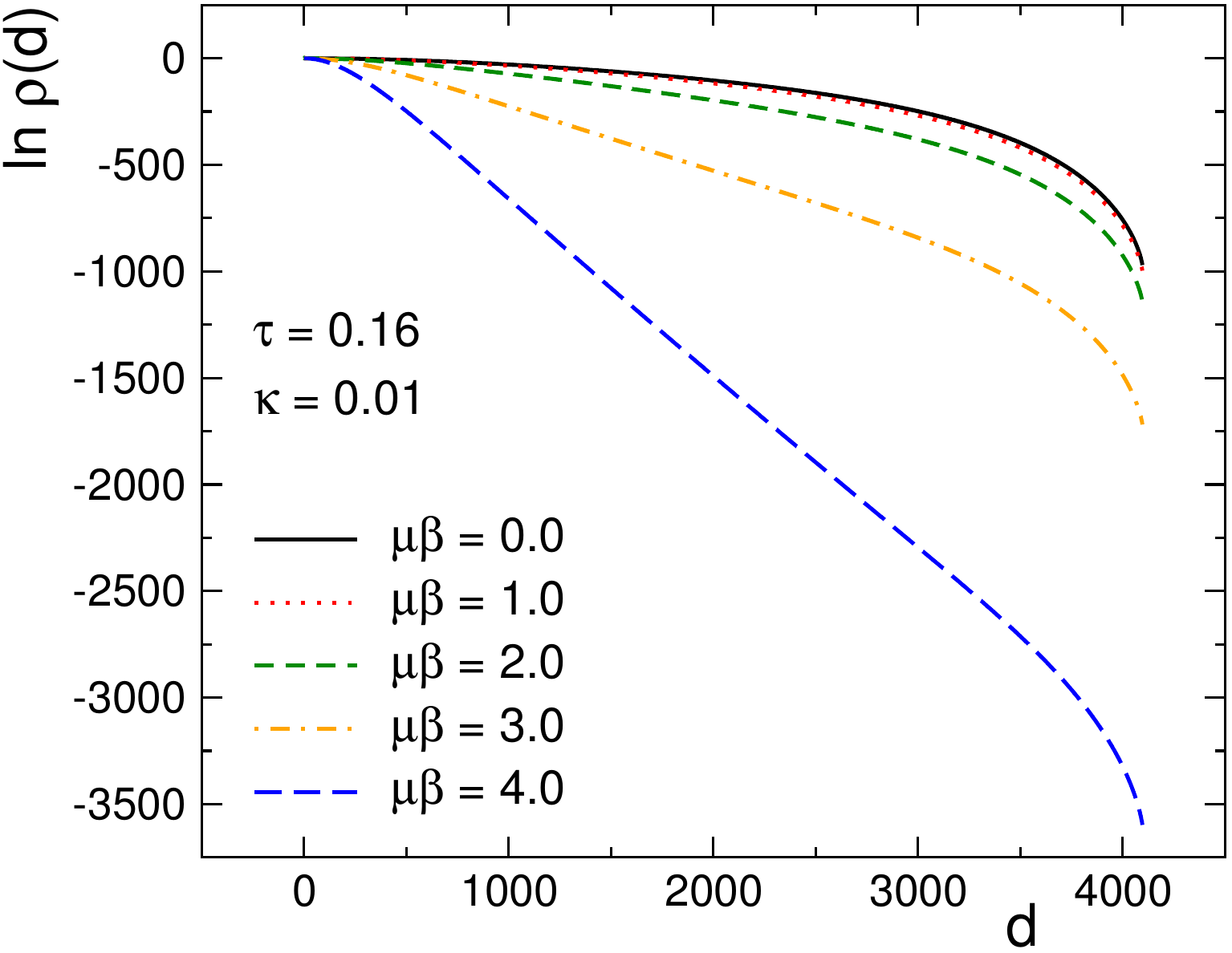}
\hspace{5mm}
\includegraphics[height=56mm]{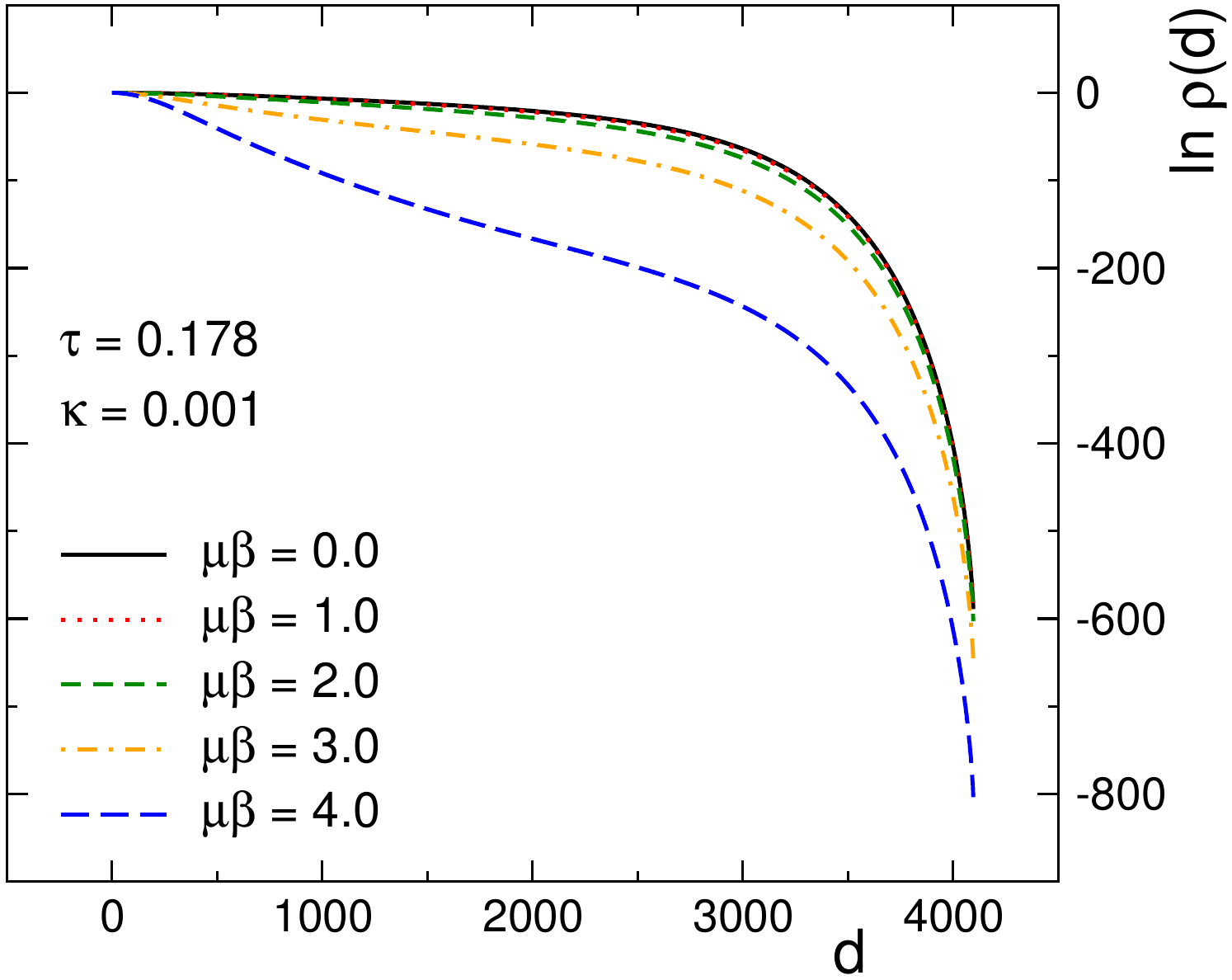}
\caption{Results for the logarithm of $\rho(d)$ as a function of $d$ from a $16^3$ lattice for different values of $\mu\beta$. 
The data we show in the lhs.\ plot are for 
$\tau = 0.16$, $\kappa = 0.01$. On the rhs.\ we use $\tau = 0.178$ and $\kappa = 0.001$. 
The error bars are smaller than the line-width. 
Note the different vertical scales for the two plots.}
\label{rhoexample}	
\end{figure}

\vskip3mm
    
\noindent  
{\bf Restricted Monte Carlo:} 
\vskip2mm

\noindent 
The FFA variant of the DoS method described here is based on fitting the Monte Carlo data for the restricted 
expectation values $\langle \langle \Delta N \rangle \rangle_n(\lambda)$ as defined in (\ref{vevrestricted}). For this purpose we 
first need to generate an initial configuration $P$ of the spin variables, such that the constraint $\Delta N [P] \in \{ n- 1, n, n +1\}$ 
is obeyed\footnote{This is the constraint for $n > 0$. The modification to the $n = 0$ case is trivial and we 
omit the discussion of this special case.}. Such a configuration can easily be constructed by hand, but of course needs to be 
equilibrated before taking measurements. For this and the subsequent computation of observables a slightly modified Monte Carlo update 
can be used. It contains an additional restriction which rejects trial configurations that violate the constraint 
$\Delta N [P] \in \{ n- 1, n, n +1\}$. The acceptance rate is very good throughout and only for $n$ very close to the
maximum value of $n=V$ (i.e., the cases $n = V-2, V -1, V$) we observe a drop in the acceptance rate. In principle it is easy to 
compute $\rho(d)$ for these largest values of $d$ exactly with a low temperature expansion. However, since for the values of $d$ 
where the quality of the Monte Carlo data decreases $\rho(d)$ is already very small, we simply use the data as we obtain them from
the simulation. In this letter we show results for lattice volumes of $10^3$ and $16^3$ and in both cases 
used $10^6$ equilibration sweeps and a statistics of $10^6$ measurements separated 
by 100 sweeps for decorrelation, and all errors we display are statistical errors.

In Fig.~\ref{fig_coeffs} we show the Monte Carlo results (symbols) for 
$\langle \langle \Delta N \rangle \rangle_n(\lambda) - n$ with $n = 0$, 2, 10, 50, 100, 300, 600, 900, 950, 990, 995 and 998  
for several values of
$\lambda$ in the interval $[-7,7]$. The data were generated on $10^3$ lattices for the parameter values $\tau = 0.16$, 
$\kappa = 0.01$ and $\mu \beta= 1.0$. 
The figure demonstrates that the Monte Carlo data show the expected simple behavior as a function of $\lambda$ and can easily 
be fit (we use a standard $\chi^2$ procedure) with the functions given in the right hand sides of (\ref{vevn}).
In Fig.~\ref{fig_coeffs} the results of the fits are shown as full curves and obviously describe the numerical data very well.
 
Once the coefficients $a_j$ are determined from the fits, we can build up the density of states $\rho(d)$ as given in 
(\ref{rhoparam}).  Results for the density $\rho(d)$ at different values of $\mu \beta$ are presented in Fig.~\ref{rhoexample}.  
The data we show in the lhs.\ plot are for $16^3$ lattices with
$\tau = 0.16$, $\kappa = 0.01$, while in the rhs.~plot $\tau = 0.178$, $\kappa = 0.001$ were used. It is remarkable that the range 
of the values for $\rho(d)$ strongly depends on the parameters, including also $\mu \beta$. This is due to the fact that the weighted density 
we use here also includes the Boltzmann factor $e^{S_R}$.

\begin{figure}[t]
\centering
\hspace*{-2.2mm}
\includegraphics[height=52.2mm]{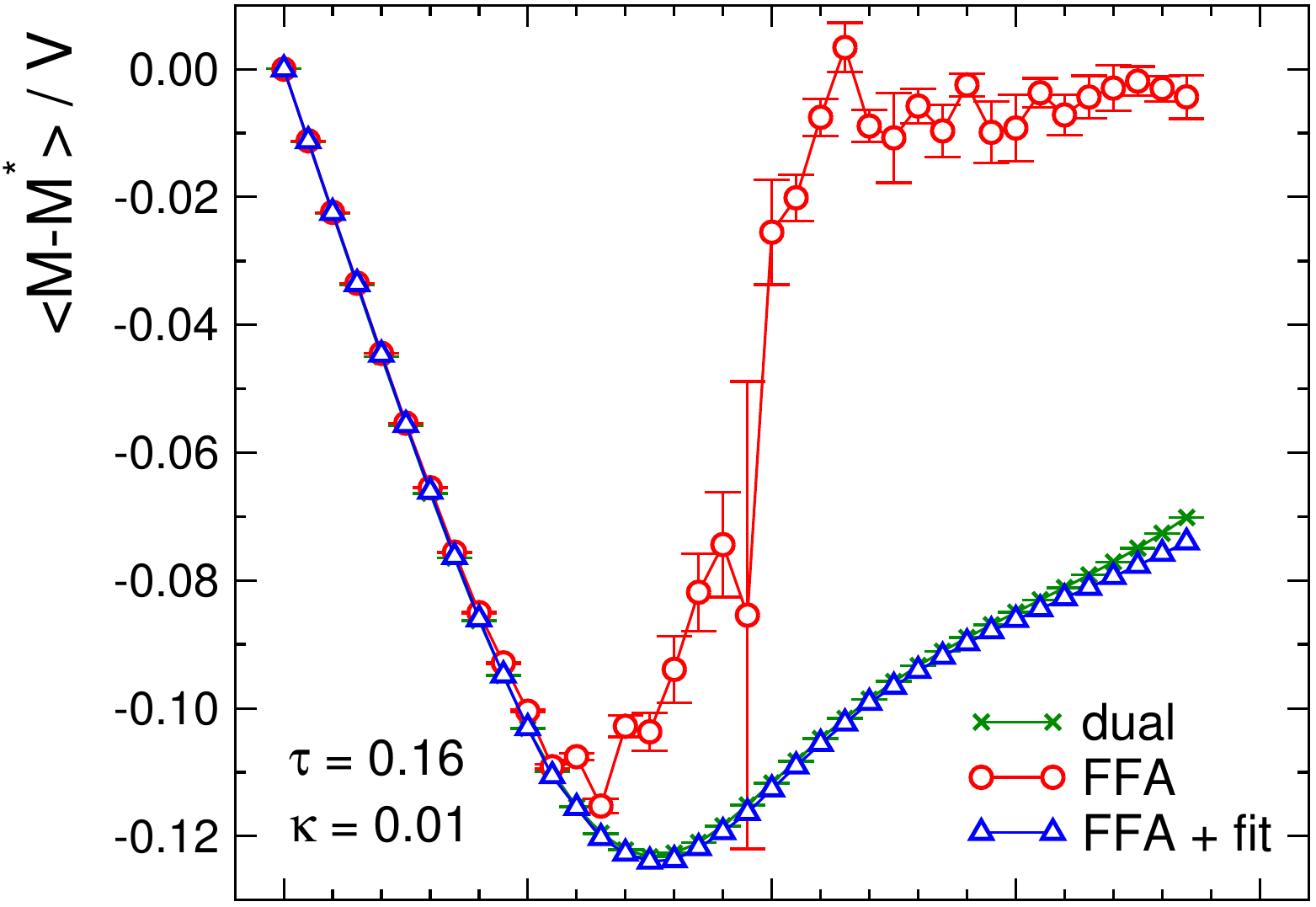}
\hspace{3.1mm}
\includegraphics[height=52.2mm]{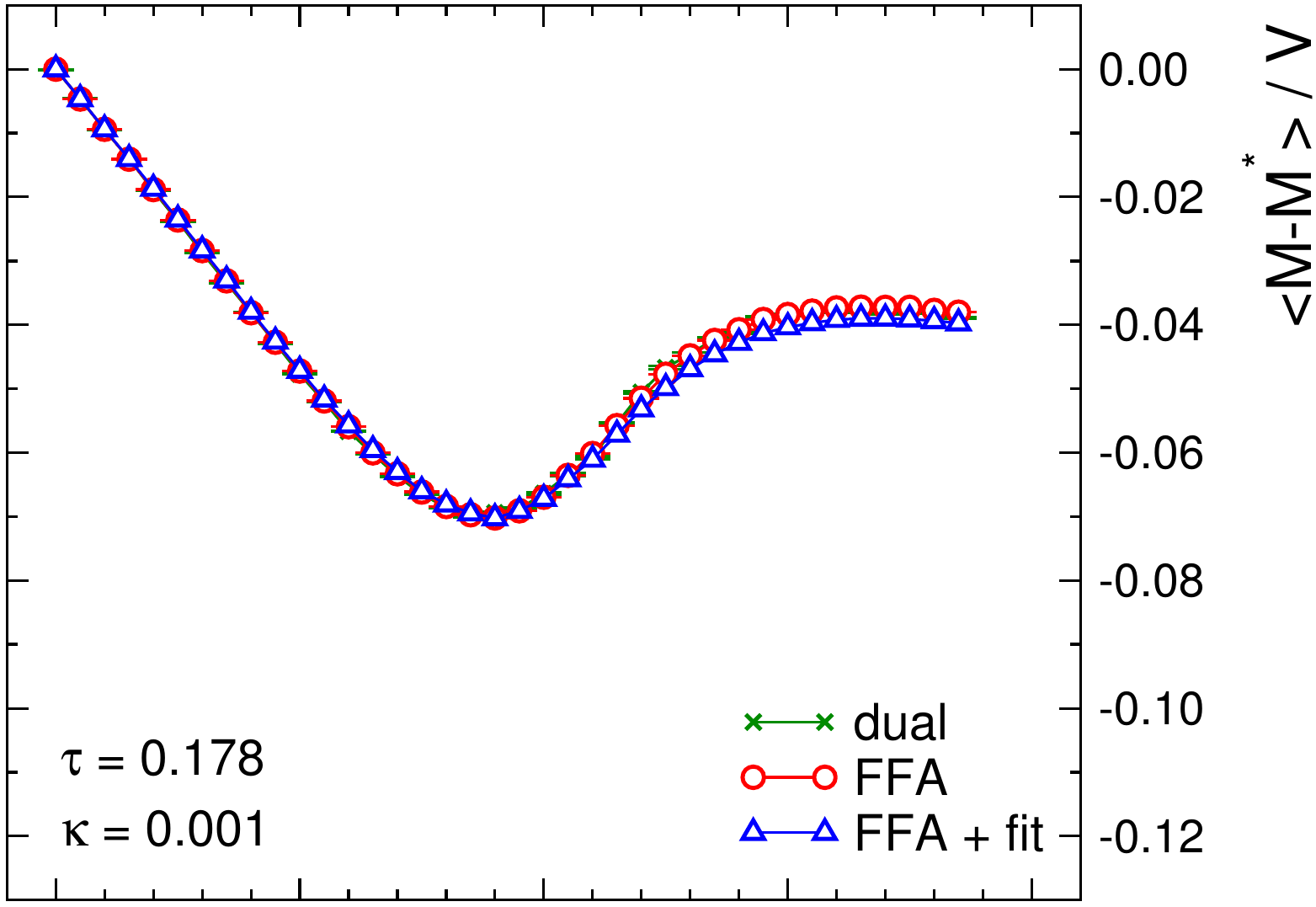}
\vspace{5mm}
\hspace*{-2mm}
\includegraphics[height=58mm]{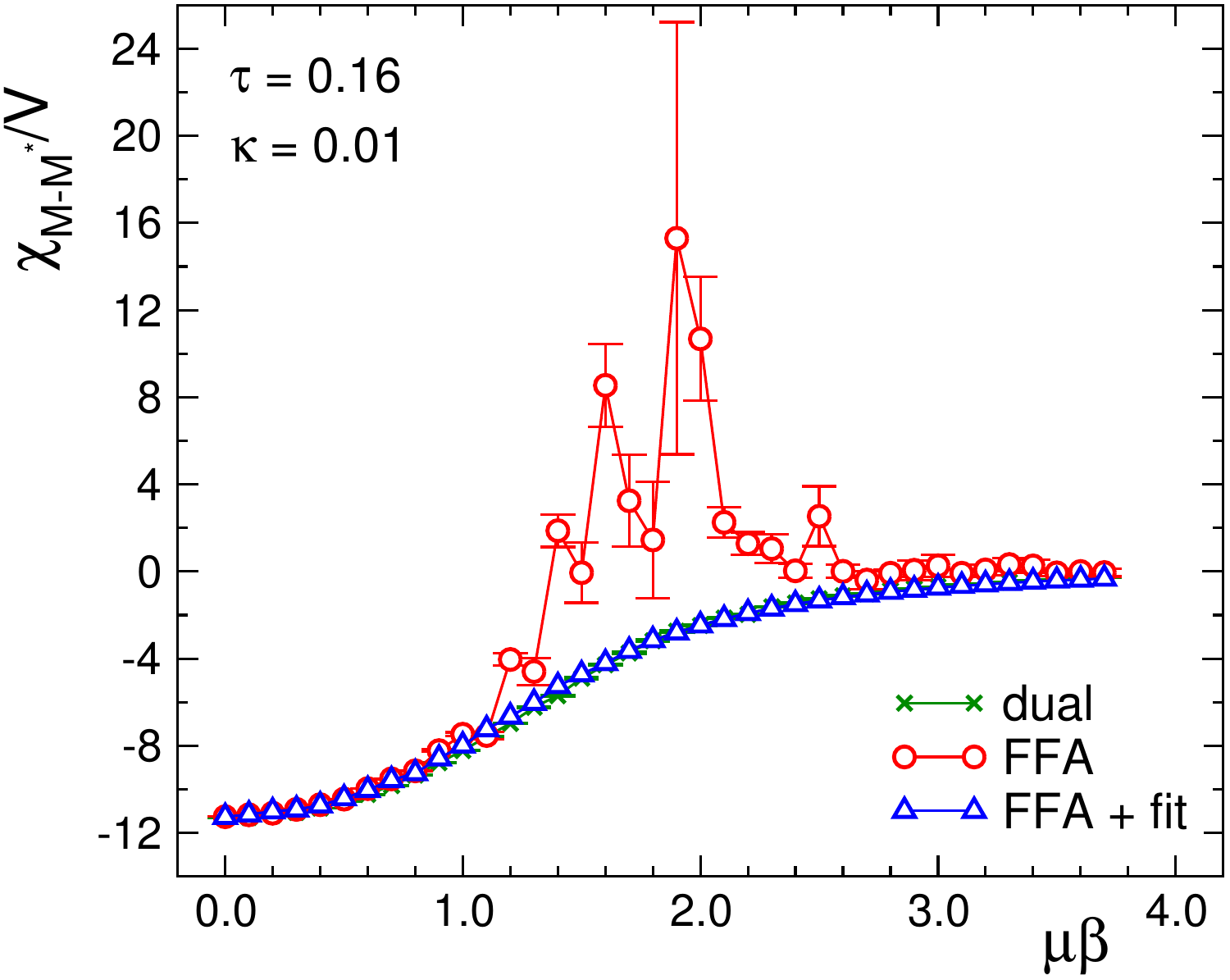}
\hspace{3mm}
\includegraphics[height=58mm]{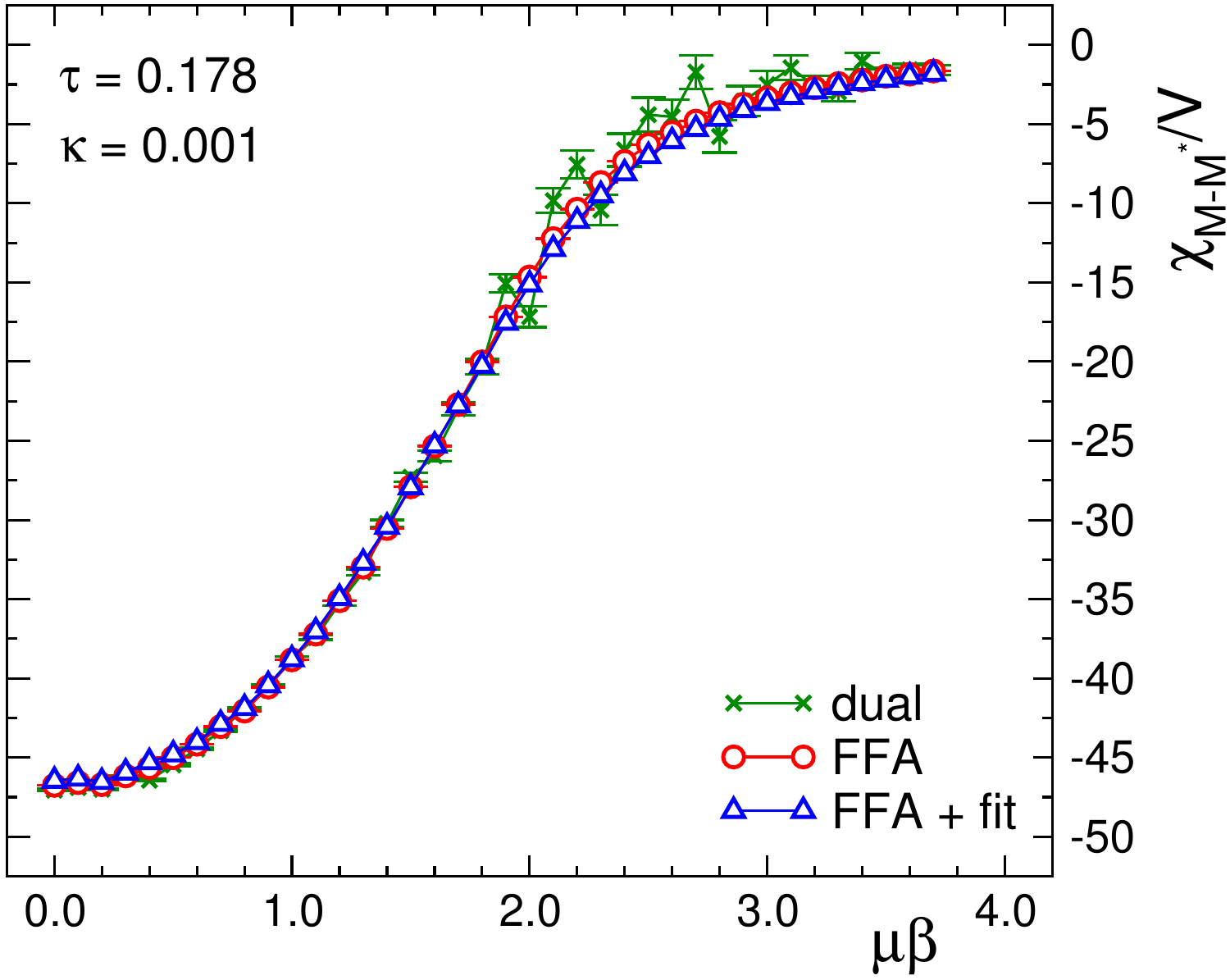}
\caption{Results for the physical observables $\langle M-M^*\rangle$ (top row of plots) and $\chi_{M-M^*}$ (bottom) on $16^3$ lattices
as a function of $\mu \beta$. We use  
two sets of parameters, $\tau = 0.16$, $\kappa = 0.01$ on the lhs., and $\tau = 0.178$ and $\kappa = 0.001$ on the rhs. 
We show results from the FFA algorithm, the FFA algorithm combined with a fit of $\rho(d)$ and for comparison also the 
results from a simulation in the dual representation.}
\label{observables}	
\end{figure}

\vskip3mm
    
\noindent  
{\bf Results for physical observables:} 
\vskip2mm

\noindent 
Having determined the density $\rho(d)$ we can finalize the calculation and evaluate the expectation values of observables
using (\ref{vev}). We consider 
$\langle M - M^* \rangle$  and $\chi_{M - M^*} =  \langle (M - M^*)^2 \rangle - \langle M - M^* \rangle^2$,
where
\begin{equation}
M \; = \; \sum_x P_x \; = \; N_0[P] \, - \, \frac{1}{2} \big(N_+[P] + N_-[P] \big) + i\frac{\sqrt{3}}{2} \big(N_+[P] - N_-[P]\big) \; ,
\end{equation}
such that $M - M^* \; = \; i \sqrt{3} (N_+[P] - N_-[P]) \; = \; i \sqrt{3} d$.
Thus for the evaluation of $\langle M - M^* \rangle$ only an odd part appears in (\ref{vev}) and thus this expectation value is real. 

In Fig.~\ref{observables} we show the results from the FFA for $\langle M - M^* \rangle$ and $\chi_{M - M^*}$ on a
$16^3$ lattice as a function of 
$\mu\beta$ (red circles). 
We display results for two sets of parameters, $\tau = 0.16$, $\kappa = 0.01$ on the lhs., and $\tau = 0.178$, $\kappa = 0.001$ (rhs.). 
The data are for $16^3$ lattices and the statistics is the same as used for the density discussed above. 
As reference data in Fig.~\ref{observables} we also show the results from a dual simulation. 
For the $\tau = 0.178, \kappa = 0.001$ data (rhs.) the FFA results agree well with the dual results for all values of $\mu\beta$ 
we show. 
However, for $\tau = 0.16$, $\kappa = 0.01$ (lhs.) the dual and the plain FFA data disagree for $\mu\beta$ larger than 2.
This discrepancy can be attributed to the fact, that for this parameter set the sign problem is much harder than for $\tau = 0.178, \kappa = 0.001$,
as can, e.g., be seen in plots for the expectation value of the phase of the action (Fig.~1 of \cite{z3fugexp}). 

The numerical problems can be related to small statistical fluctuations of $\rho(d)$ around its exact values \cite{doskurt2}. 
However, it is straightforward to 
smoothen these local fluctuations by using a fit of $\rho(d)$ and then computing the observables (\ref{vev}) with the fitted $\rho(d)$. For the fit 
we use a polynomial in $d^2$ (note that $\rho(d)$ is an even function and $\ln \rho(0)=0$) for the logarithm of $\rho(d)$, i.e.,
$\ln \rho(d) =  \sum_{n=1}^N \, c_n \, d^{\,2n}$. 
In Fig.~\ref{observables} we also display the results obtained from the fitted $\rho(d)$ with $N = 15$ (blue triangles) and find that we obtain a much 
larger range in $\mu\beta$ where FFA and dual simulation agree also for the $\tau = 0.16$, $\kappa = 0.01$ data. 

We stress that using a fit to smoothen $\rho(d)$ is of course not a fundamental ingredient of the 
FFA, and increasing the statistics when determining $\rho(d)$ will also improve the results. 
However, the source of the error is very clear: For large $\mu\beta$ the density $\rho(d)$ is probed by the rapidly 
oscillating factors in (\ref{zfinal}) and (\ref{vev}) and the small fluctuations in $\rho(d)$ become
dominant. On the other hand $\rho(d)$ is a smooth function, such that a fit is a much more cost efficient method than a drastic increase of 
the statistics.

\begin{figure}[t]
\centering
\hspace*{-5mm}
\includegraphics[height=63mm]{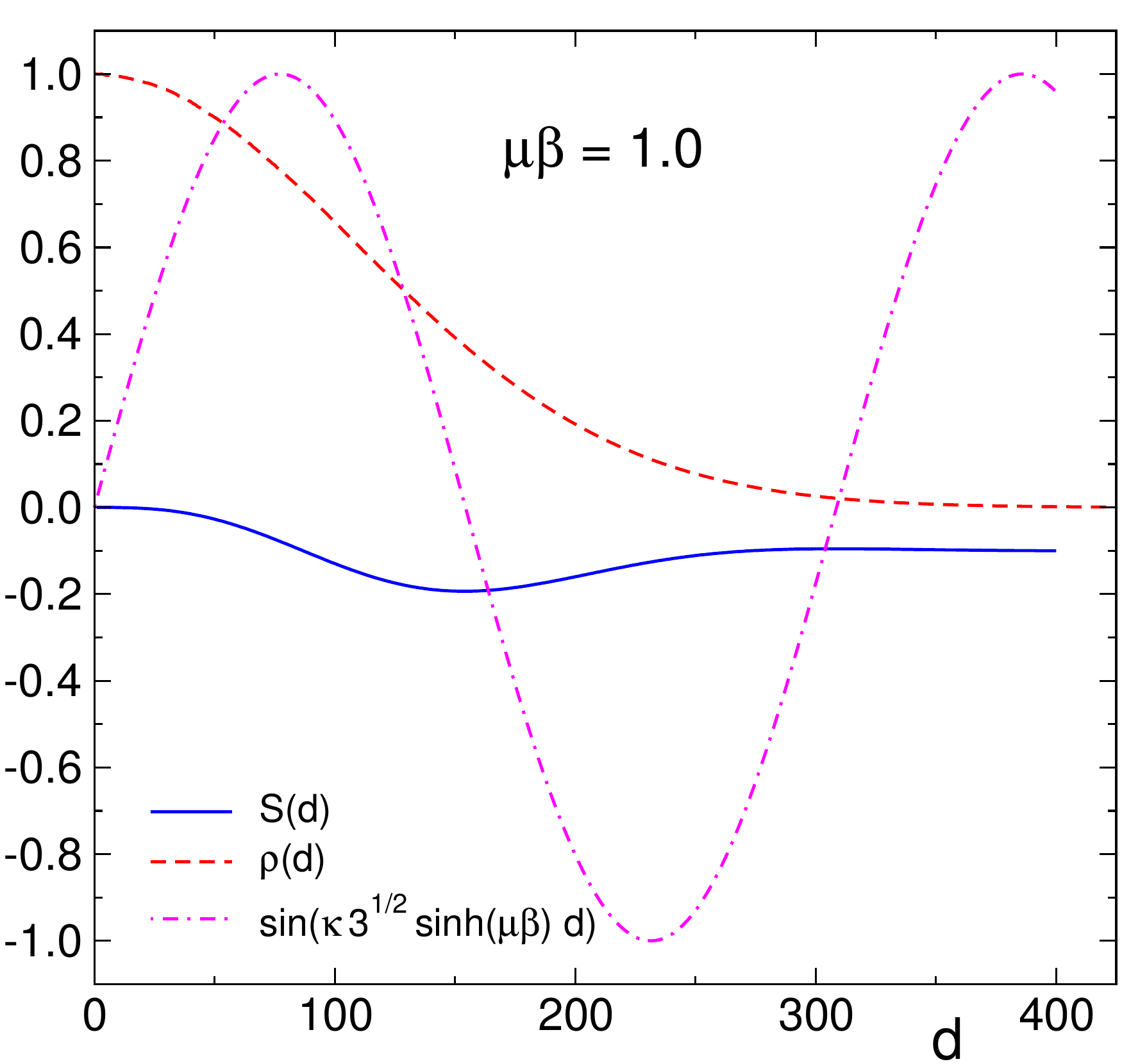}
\hspace{3mm}
\includegraphics[height=63mm]{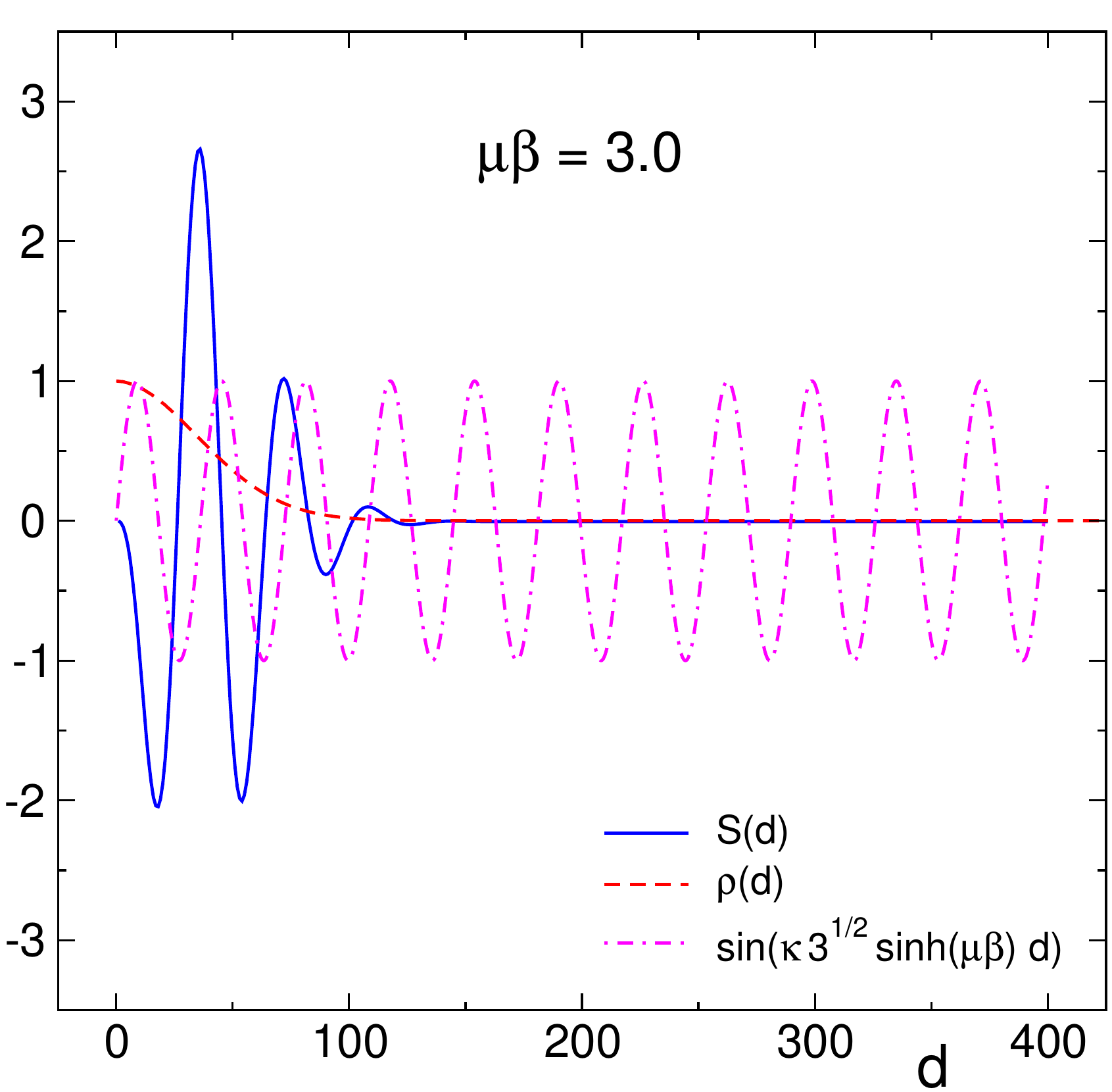}
\caption{Buildup of the contributions to $\langle M - M^* \rangle/V$ for  $V = 16^3$, $\tau = 0.16$ and $\kappa = 0.01$ at $\mu \beta = 1.0$ (lhs.) 
and $\mu \beta= 3.0$ (rhs.). See the text for a discussion of the plots.}
\label{comsum}	
\vskip6mm
\centering
\hspace*{-0.7mm}
\includegraphics[height=58mm]{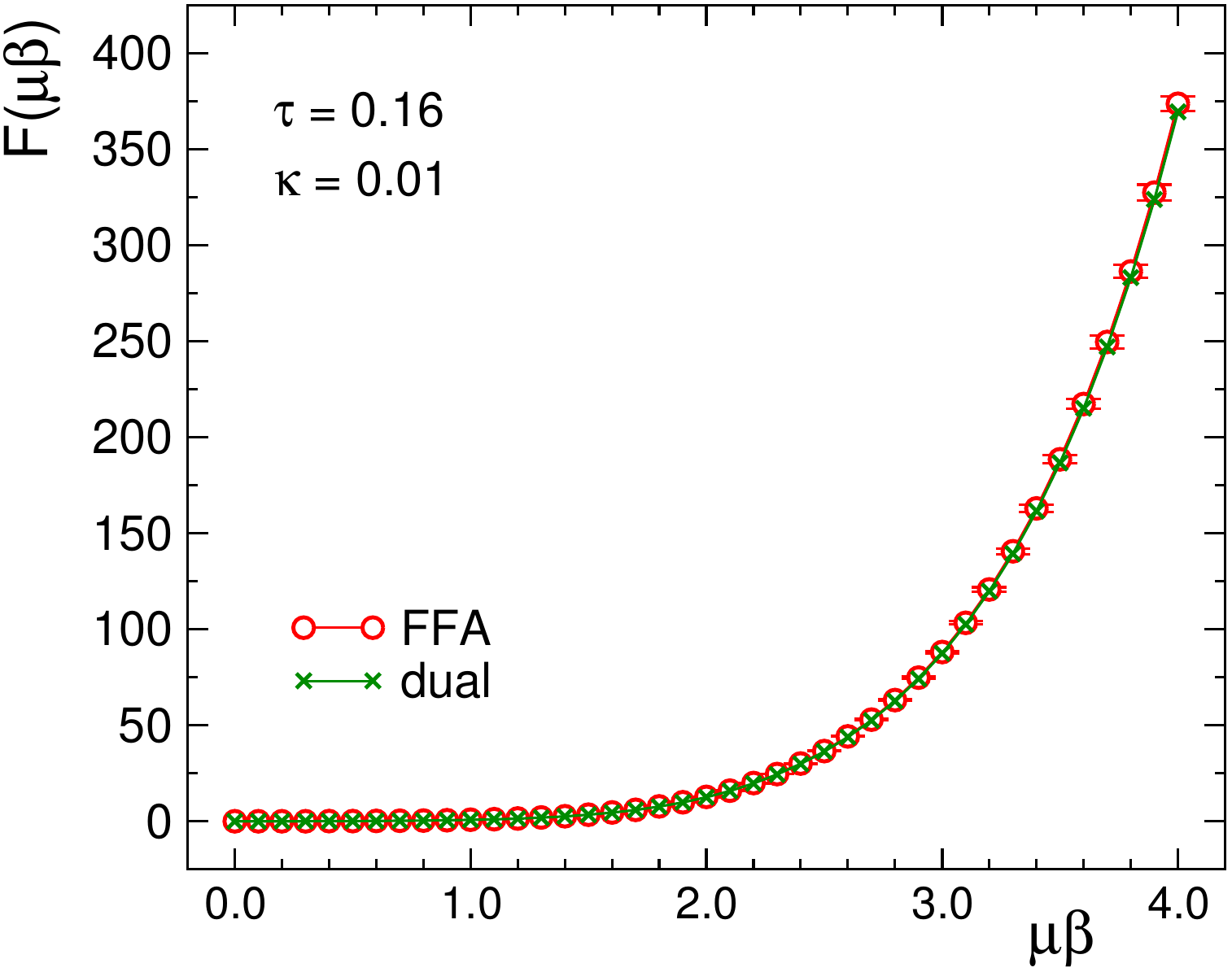}
\hspace{5mm}
\includegraphics[height=58mm]{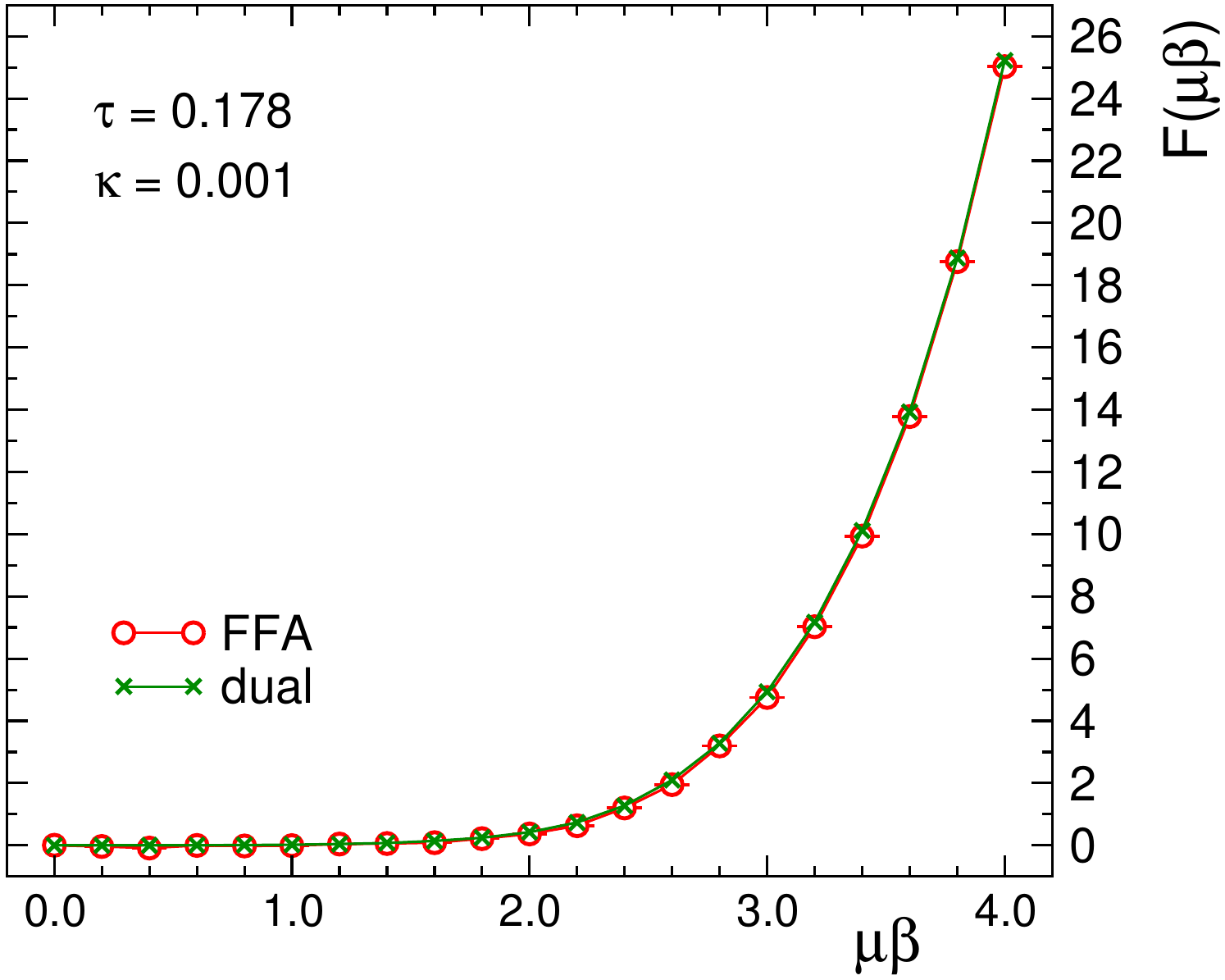}
\caption{Results for the the free energy $F(\mu\beta)$ from the FFA method (circles) and from the dual simulation (crosses). 
The data are for $V = 10^3$ with $\tau = 0.16$, $\kappa = 0.01$ (lhs.), and $\tau=0.178$, $\kappa = 0.001$ (rhs.).}
\label{fe}	
\vspace{-3mm}
\end{figure}

The influence of small fluctuations of the density $\rho(d)$ is studied in Fig.~\ref{comsum}. In this plot we zoom into the lower values of $d$ 
(note that $d$ runs from 0 to $V = 16^3 = 4096$), and show as a function of $d$ the density 
$\rho(d)$, the oscillating factor $\sin(\kappa \sqrt{3} \sinh(\mu \beta)d)$ and the cumulative sum 
$S(d) = -  2 \sqrt{3} /VZ \sum_{j = 1}^d \rho(j) \sin(\kappa \sqrt{3} \sinh(\mu \beta) j) j$, which for $d = V$ sums up to 
$\langle M - M^* \rangle/V$. Thus studying the cumulative sum $S(d)$ as a function 
of $d$ shows how $\langle M - M^* \rangle/V$ is built up.

For the smaller value $\mu \beta = 1.0$ shown in the lhs.~plot, we find that $S(d)$ 
shows no sizable fluctuations and approaches its asymptotic value 
without major fluctuations. For $\mu \beta = 3.0$, however, we observe that $S(d)$ strongly picks up the now faster oscillations from 
$\sin(\kappa \sqrt{3} \sinh(\mu \beta)d)$ and crosses the value $S(d)$ several times. This implies that large cancellations are necessary to reach 
the asymptotic value and small fluctuations of the density $\rho(d)$ have a large impact. These fluctuations are suppressed when fitting the
density as discussed above.

As our final observable we consider the free energy $F(\mu \beta)$ defined as
$F(\mu\beta) = \ln ( Z(\mu \beta) / Z(0) )$. In Fig.~\ref{fe} we show the results for the free energy at two values of the couplings
and again compare the FFA method with data obtained from a dual simulation. As for the other observables we find 
excellent agreement of the FFA results with the data from the reference simulation in the dual approach. It is remarkable that for the
free energy this excellent agreement is achieved already without fitting the density, which was necessary for the bulk observables. 
    
\noindent  
{\bf Summary:} 
\vskip2mm

\noindent 
The key issue in the application of the density of states method is to determine 
the density $\rho(d)$ as precise as possible, since in the expectation values (\ref{vev}) $\rho(d)$ is summed over with a highly
oscillating function. Thus it is necessary to optimize the strategy for the computation of $\rho(d)$ in every possible way. 
In these notes we test a strategy (the functional fit approach (FFA)) where in restricted Monte Carlo simulations on small intervals on $d$
the response of the system to a free parameter in the Boltzmann factor is evaluated. The response is given by a known function which we
fit to the Monte Carlo data to determine the parameters of the density $\rho(d)$. 

The results from the DoS calculation for $\langle M - M^* \rangle$, $\chi_{M - M^*}$ and the free energy $F(\mu \beta)$ are compared 
to reference data obtained in a dual simulation. We show that when using a fit for the density of states $\rho(d)$ the observables agree very
well with the dual results on a surprisingly large range of $\mu \beta$ values up to $\mu \beta \approx 4$.

\vskip2mm
\noindent  
{\bf Acknowledgements:} We thank Biagio Lucini and Kurt Langfeld for interesting discussions. Pascal T\"orek is supported 
by the FWF Doktoratskolleg DK W 1203 ''Hadrons in Vacuum, Nuclei and Stars''. Furthermore this work is partly supported by 
the Austrian Science Fund FWF Grant. Nr. I 1452-N27 and by DFG TR55, ''Hadron 
Properties from Lattice QCD''.


\begin{thebibliography}{1234567}

\bibitem{dosold}
   A.~Gocksch,
  Phys.\ Rev.\ Lett.\  {\bf 61} (1988) 2054.

\bibitem{dosrecent1}
Z.~Fodor, S.D.~Katz and C.~Schmidt,
  JHEP {\bf 0703} (2007) 121
  [hep-lat/0701022].
%
  C.~Schmidt, Z.~Fodor and S.D.~Katz,
  PoS LAT {\bf 2005} (2006) 163
  [hep-lat/0510087].

\bibitem{dosrecent2}
  S.~Ejiri,
  Phys.\ Rev.\ D {\bf 77} (2008) 014508
  [arXiv:0706.3549].
%
  Y.~Nakagawa {\it et al.}  [WHOT-QCD Collaboration],
  PoS LATTICE {\bf 2011} (2011) 208
  [arXiv:1111.2116].
%
  S.~Ejiri {\it et al.}  [WHOT-QCD Collaboration],
  Central Eur.\ J.\ Phys.\  {\bf 10} (2012) 1322
  [arXiv:1203.3793];
%
  PoS LATTICE {\bf 2012} (2012) 089
  [arXiv:1212.0762].
%
  S.~Ejiri,
  Eur.\ Phys.\ J.\ A {\bf 49} (2013) 86
  [arXiv:1306.0295].
%
J.~Greensite, J.C.~Myers and K.~Splittorff,
  JHEP {\bf 1310} (2013) 192
  [arXiv:1308.6712].

\bibitem{doskurt1}
  K.~Langfeld, B.~Lucini, A.~Rago,
  Phys.\ Rev.\ Lett.\  {\bf 109} (2012) 111601
  [arXiv:1204.3243].
%
   K.~Langfeld, J.~Pawlowski, B.~Lucini, A.~Rago, R.~Pellegrini,
  PoS LATTICE {\bf 2013} (2014) 198
  [arXiv:1310.8231].
%
  R.~Pellegrini, K.~Langfeld, B.~Lucini, A.~Rago,
  PoS LATTICE {\bf 2014}
  [arXiv:1411.0655].
 
\bibitem{wanglandau}  
F. Wang and D.P. Landau,
Phys. Rev. Lett. 86 (2001) 2050.  
   
\bibitem{doskurt2}
%
  K.~Langfeld and B.~Lucini,
  Phys.\ Rev.\ D {\bf 90} (2014) 9,  094502
  [arXiv:1404.7187].
%
  B.~Lucini and K.~Langfeld,
  PoS LATTICE {\bf 2014}
  arXiv:1411.0174.
%
  K.~Langfeld, B.~Lucini, A.~Rago, R.~Pellegrini and L.~Bongiovanni,
  arXiv:1503.00450.

\bibitem{z3model}
  A.~Patel, Nucl.~Phys. B {\bf 243} (1984) 411;
  Phys.\ Lett.\  B {\bf 139} (1984) 394.
%
  T.~DeGrand, C.~DeTar,  
  Nucl.\ Phys.\  B {\bf 225} (1983) 590.    
%
  J.~Condella, C.E.~DeTar,
  Phys.\ Rev.\  D {\bf 61} (2000) 074023.
%
  M.G.~Alford, S.~Chandrasekharan, J.~Cox, U.-J.~Wiese,
  Nucl.\ Phys.\  B {\bf 602} (2001) 61.
%
  S.~Kim, P.~de Forcrand, S.~Kratochvila, T.~Takaishi,
  PoS {\bf LAT2005} (2006) 166.

\bibitem{z3dual}
  Y.D.~Mercado, H.G.~Evertz and C.~Gattringer,
  Phys.\ Rev.\ Lett.\  {\bf 106} (2011) 222001
  [arXiv:1102.3096];
%
  Comput.\ Phys.\ Commun.\  {\bf 183} (2012) 1920
  [arXiv:1202.4293].

\bibitem{poslat}
  Y.D.~Mercado, P.~T\"orek and C.~Gattringer,
  PoS(LATTICE2014)203 [arXiv:1410.1645].
  
\bibitem{effmod}
L.G.~Yaffe,~B.~Svetitsky,  Nucl.~Phys.~B {\bf 210} 423. 
  
\bibitem{z3fugexp}
  E.~Gr\"unwald, Y.D.~Mercado and C.~Gattringer,
  Int.\ J.\ Mod.\ Phys.\ A {\bf 29} (2014) 32, 1450198
  [arXiv:1403.2086];
%
  PoS LATTICE {\bf 2013} (2014) 448
  [arXiv:1310.6520].
       
\end{thebibliography}
\end{document}